# Building XenoBuntu Linux Distribution for Teaching and Prototyping Real-Time Operating Systems

Nabil LITAYEM, Ahmed BEN ACHBALLAH, Slim BEN SAOUD

Department of Electrical Engineering - INSAT,

University of Carthage, TUNISIA

{nabil.litayem, ahmed.achballah, slim.bensaoud}@gmail.com

*Abstract-* **This paper describes the realization of a new Linux distribution based on Ubuntu Linux and Xenomai Real-Time framework. This realization is motivated by the eminent need of real-time systems in modern computer science courses. The majority of the technical choices are made after qualitative comparison. The main goal of this distribution is to offer standard Operating Systems (OS) that include Xenomai infrastructure and the essential tools to begin hard real-time application development inside a convivial desktop environment. The released live/installable DVD can be adopted to emulate several classic RTOS Application Program Interfaces (APIs), directly use and understand real-time Linux in convivial desktop environment and prototyping real-time embedded applications.**

## I. INTRODUCTION

Real-Time embedded software become an important part of the information technology market. This kind of technology previously reserved to very small set of mission-critical applications like space crafts and avionics, is actually present in most of the current electronic usage devices such as cell phones, PDAs, sensor nodes and other embedded-control systems [1]. These facts make the familiarization of graduate students with embedded real-time operating systems very important [2]. However, many academic computer science programs focus on PC based courses with proprietary operating systems. This could be interesting for professional training, but inappropriate to the academicians because it limits students to these proprietary solutions.

In the RTOS market, there are some predominant actors with industry-adopted standards. Academic real-time systems courses must offer to the students the opportunity to use and understand the most common RTOSs APIs. Actually, we assist to the growing interest of the real-time Linux extensions. In fact, they must be considered with great interests since each real-time Linux extension offers a set of advantages [3]. Xenomai real-time Linux extensions have the main advantages to emulate standard RTOS interfaces, compatibility with non-real-time Linux. Such adoption can be very cost-effective for the overall system.

In this paper, we present the interest of using Xenomai and Ubuntu as live installable DVD for teaching real-time operating systems and rapid real-time applications prototyping. Technical choices and benefits of the chosen solutions will be discussed.

The remainder of this paper is organized as follows. Section 2 presents a survey of RTOS market and discusses both of the classic solution and the Linux-based alternatives. The Remastering solutions and available tools are detailed in Section 3. Section 4 describes the realization of our live DVD. Conclusions and discussion are provided in Section 5.

## II. SURVEY OF THE RTOS MARKET

### A. Classic RTOS and Real-Time API

RTOS is an essential building block of many embedded systems. The most basic purpose of RTOS is to offer task deadline handling in addition to classic operating system functionalities. The RTOS market is shared between few actors. Each of them has its appropriate development tools, its supported target, its compiler tool-chain and its RTOS APIs. In addition, several RTOS vendors can offer additional services such as protocol stacks and application domain certification.

According to the wide varieties of RTOSs, the designers must choose the most suitable one for their application domain. In the following, we will present a brief description of traditional RTOS and real-time API available in the embedded market.

#### 1) VxWorks

VxWorks [4] is a RTOS made and sold by Wind River Systems actually acquired by Intel. It was primary designed for embedded systems use. VxWorks continues to be considered as the reference RTOS due to its wide range of supported targets and the quality of its associated IDE.

#### 2) PSOS

This RTOS [5] was created in about 1982. It was widely adopted especially for Motorola MCU. Since 1999 PSOS has been acquired by Wind River Systems.

*3)* *VRTX*

VRTX [6] is an RTOS suitable for both traditional board-based embedded systems and **S**ystem **o**n **C**hip (SoC). It was widely adopted for RISC microprocessors.

*4)* *POSIX*

POSIX [7] (**P**ortable **O**perating **S**ystem **I**nterface for **C**omputer **E**nvironments), is a set of a standardized interface that provides source level compliance for RTOS services.

### B. Real Time Linux Alternatives

According to the reference study [4], the market place of embedded Linux becomes more and more important. In 2007 their part was about 47% of the total embedded market. The same study anticipates that the market place of embedded Linux will be 70% in 2012. These facts can be justified by the growing availability resources in modern embedded hardware, the maturity of actual Linux kernels and applications and the cost reduction needs. Actually, there are many existing open source implementations of real-time extensions for Linux kernel, but we must note that various existing industrial solutions are based on those extensions with an additional value of support quality. Real-time Linux variants are actually successfully used in different applications [5]. Due to the increasing Linux popularity in the embedded systems' field, many efforts were spent and proposed to transform Linux kernel into a real-time solution. These works resulted in several implementations of real-time Linux. Actually, there are many existing implementations of real-time extension for Linux kernelextension [6]. They can be classified in two categories according to the approach used to improve their real-time performance of the Linux kernel. The first approach consists of modifying the kernel behavior to improve its real-time characteristics. The second approach consists of using a small real-time kernel to handle real-time tasks and what can run the Linux kernel as a low priority task.

Actually, a lot of researches and industrial efforts are made to enhance the real-time capability of the various real-time Linux flavors' [3] for different perspectives and applications domain. These works can be classified in two categories. The first one is about scheduling algorithm and timer management. The second category is about application's domain such as Hardware-in-the-Loop simulation system, model based engineering [7] and real-time simulation. In Table I, we present some of the available open source many research Linux implementations.

TABLE I.        LINUX OPEN SOURCE RTOSS

| Linux-based RTOS | Description |
|---|---|
| ADEOS | **A**daptive **D**omain **E**nvironment for **O**perating **S**ystems) [11], is a GPL nanokernel hardware abstraction layer created to provide a flexible environment for sharing hardware resources among many operating systems. *ADEOS* enables multiple prioritized domains to exist simultaneously on the same hardware. |
| ART Linux | **A**dvanced **R**eal-**T**ime Linux [12], is a hard real-time Linux extension inspired from RTLinux and developed with robotics applications in mind. Real-Time is accessible from user level and does not require special device drivers. ART Linux is available for 2.2 and 2.6 Linux kernel. |
| KURT | **K**ansas **U**niversity's **R**eal-**T**ime Linux is a real-time Linux [13] extension developed by the Kansas University for x86 platforms. It can allow scheduling of events with a 10µs resolution. |
| QLinux | QLinux [14] real-time Linux kernel, is a Linux extension that focus and provide Quality of Service (QoS) guarantees for "soft real-time" performance in applications such as multimedia, data collection, etc. |
| Linux/RK | **L**inux **R**esource **K**ernel [15] is a real-time extension which incorporates real-time services to the Linux kernel. |
| RTAI | **R**eal-**T**ime **A**pplication [16] **I**nterface usable both for mono processors and symmetric multi-processors (SMPs), that allows the use of Linux in many "hard real-time" applications. RTAI is the real-time *Linux* that has the best integration with other open source tools scilab/scicos and Comedi. This extension is widely used in control applications. |
| Xenomai | *Xenomai* [17] is a real-time development framework that provides hard real-time support for GNU/Linux. It implements *ADEOS* (I-Pipe) micro-kernel between the hardware and the Linux kernel. I-Pipe is responsible for executing real-time tasks and intercepts interrupts, blocking them from reaching the Linux kernel to prevent the preemption of real-time tasks by Linux kernel. *Xenomai* provides real-time interfaces either to kernel-space modules or to user-space applications. Interfaces include RTOS interfaces (*pSOS+*, *VRTX*, *VxWorks*, and *RTAI*), standardized interfaces (*POSIX*, *uITRON*), or new interfaces designed with the help of *RTAI* (native interface).**T**he**s**e features made that *Xenomai* was considered as the RTOS Chameleon for Linux. It was designed for enabling smooth migration from traditional RTOS to Linux without having to rewrite the entire application. |
| RT-Preempt | The RT-Preempt patch [18] converts Linux into a fully preemptible kernel. It allows nearly the entire kernel to be preempted, except for a few very small regions of code. This is done by replacing most kernel spinlocks with mutexes that support priority inheritance and are preemptive, as well as moving all interrupts to kernel threads. (Dubbed interrupt threading), which by giving them their own context allows them to sleep among other things. |

### C. Selecting real-time extension for educational puroposes

Xenomai, RTAI and RT-Prempt are the most used real-time Linux extensions. According to the study [8], Xenomai and RTAI can provide interesting performances comparable to those offered by VxWorks in hard real-time applications. RTAI has the best integration with open source tools and can be remarkable for teaching control application. RT-Prempt has the privilege to be integrated to the mainline kernel. It offers the support of all drivers integrated into the standard kernel. Xenomai can provide the capability of emulating classic RTOS APIs with good real-time characteristics. It can be also fully compatible with RTAI. For these reasons, we focus on Xenomai to be the primary extension to integrate in our solution.

### 1) Xenomai technology and ADEOS

To make Xenomai tasks hard real-time in GNU/Linux, a real-time application interface (RTAI) co-kernel is used. It allows real-time tasks to run seamlessly aside of the hosting GNU/Linux system while the tasks of the "regular" Linux kernel are seen as running in a low-priority mode. This cohabitation is done using the previously presented ADEOS nanokernel and illustrated by figure 1.

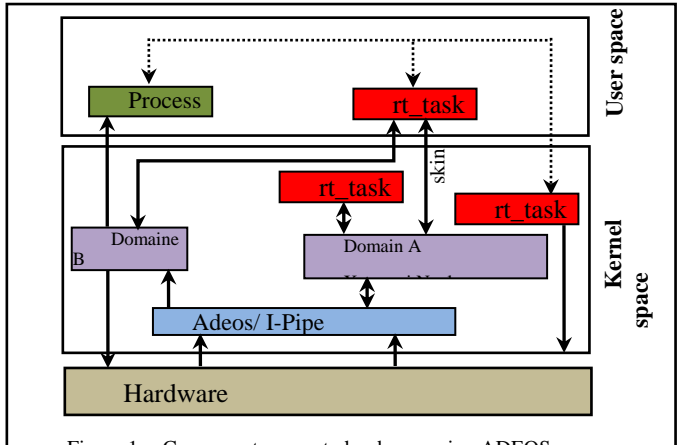

Figure 1.    Concurrent access to hardware using ADEOS

Based on the behavioral similarities between the traditional RTOS, Xenomai technology aims to provide a consistent architecture-neutral and generic emulation layer taking advantages from these similarities. This emulation can lead to fill the gap between the very fragmented RTOS world and the GNU/Linux world.

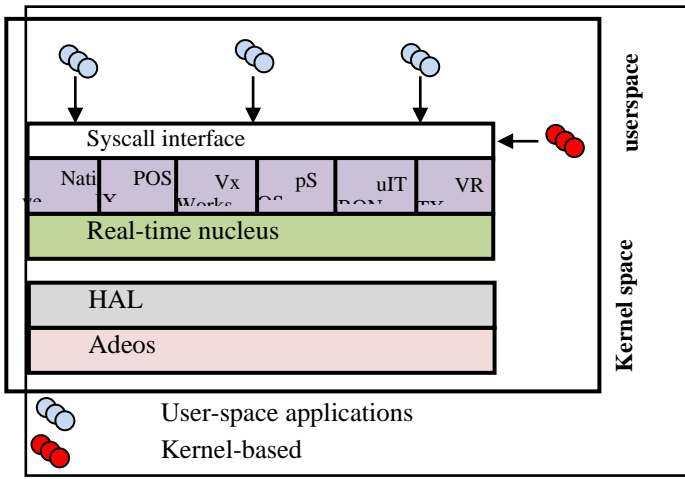

Figure 2. Xenomai Architecture

Xenomai relies on the shared features and behaviors [9] found between many embedded traditional RTOS, especially from the thread scheduling and synchronization standpoints. These similarities are used to implement a nucleus that offers a limited set of common RTOS behavior. This behavior is exhibited using services grouped in high-level interfaces that can be used in turn to implement emulation modules of real-time application programming interfaces. These interfaces can mimic the corresponding real-time kernel APIs. Xenomai technology offers a smooth and comfortable way for real-time application migration from traditional RTOS to GNU/Linux.

### 2) RTOS emulation in the industrial field

The fact that Xenomai can offer real-time capabilities in a standard desktop environment can be very useful in control system prototyping [10]. In this case, the desktop system which is running Xenomai can be used as X-in-the-loop to emulate the standard controlled equipments (electrical motors, power plants, etc.) in different phases of product prototyping and testing.

Thus, since Xenomai can emulate the most classic RTOS API, we can easily port any application developed for this RTOS to it. Furthermore, it's covered by open-source license which has a very interesting cost advantage. Moreover, by Using Xenomai we can realize an easy migration to open-source solutions without having to rewrite previously developed RT applications for proprietary RTOS. Xenomai can also reduce the application price by offering the ability to cohabit them with standard time shared Linux applications, to benefit from all the software infrastructure of Linux combined to RT capability.

### 3) RTOS emulation in academic field

Real-Time students must have a clear idea about various RTOS APIs. The cost of buying a large collection of classic RTOS to use them in the education field is not feasible. Xenomai also offers the capability of using different RTOS APIs, understanding the abstraction concept of them, manipulating the kernel/user spaces and learning about virtualization technologies.

## III.    REMASTERING UBUNTU LINUX

### A. Interest of remastering Linux distributions

A live CD or DVD allows any user to run different OS or applications without having to install them on the computer. To build a live CD/DVD, we must remaster an existing OS. Remastering is the process of customizing a software distribution. It is particularly associated with the Linux distribution world but it was extended to the majority of widely used OS. We can highlight that the most Linux distributions have been started by remastering another distribution. The term was popularized by Klaus Knopper, creator of the Knoppix Live Distribution, which has traditionally encouraged its users to modify his distribution in the way that satisfies their needs. Remastering OS can be used to make a full system backup including personal data to a live or installable CD, DVD or Flash disk that is usable and installable anywhere. It can also be exploited to make a distributable copy of an installed and customized operating system.

### B. Existing remastering software for Ubuntu Linux

There are many remastering solutions of various Linux distributions. Our live/installable DVD is based on Ubuntu because this OS has gained a growing place in different application areas. The most known remastering solutions are such as Remastersys, Ubuntu Customization Kit, Reconstructor, Builder, ulc-livecd-editor and Disc Remastering Utility.

Reconstructor and Ubuntu Customization Kit can make a personalized live system based on official image. The use of such approach is relatively complicated. The others' tools are focusing on the package installation and boot customizations. We adopted Remastersys since it is the most useful and powerful tool that we find in the available list of remastering solutions.

## IV. THE INTEGRATION OF XENOMAI IN UBUNTU INFRASTRUCTURE

Xenomai is only related to the Linux kernel version. It's independent of the Linux distribution in which it will be run. The recent Ubuntu distributions integrate Xenomai as a default package. We have taken the choice of using Ubuntu as basic distribution because it inherits all the benefit of a Debian distribution in terms of reliability and the number of available packages. Ubuntu has also the best existing Multilanguage support. Many computer constructors propose Linux as an alternative operating system. This type of systems can be used as a framework for Model-Driven Engineering (MDE) in Control and Automation since the usage of a standard operating system such as Ubuntu can facilitate the integration of these tools. The realization of our Live DVD was conducted following the steps' bellow.

### A. Adding Xenomai functionality to Linux kernel

In This step, we must firstly download the essential packages needed to configure and compile the Linux kernel. These packages are: build-essential, kernel-package, ncurses-dev. They can be installed using synaptic or apt-get. Secondly, we must download both Linux kernel and its compatible Xenomai framework, patch the Linux kernel using the prepare-kernel tool included in Xenomai package, configure, compile it and add this kernel to the boot choices. For the actual release, we used the xenomai-2.4.10 and the Linux-2.6.30.8. The compilation and installation must preferably be realized using make-kpkg tools designed especially for Debian based distributions. After realizing these steps, we can boot a system running Linux kernel using ADEOS.

### B. Compiling Xenomai and running some samples

The second step must begin by creating a Xenomai group and adding to it the appropriate users (XenoBuntu and root). We can actually configure, compile and install Xenomai and their examples, customizing available software by adding development environments. We adopted CodeLite, which is an Integrated Development Environment (IDE) designed for C and C++ development and Scilab/Scicos which can be used for control systems prototyping and real-time code generation. After rebooting our running system, we can boot to a usable system based on Xenomai through it, we can test some real-time examples based on different standards API.

### C. Transform our running system in a live DVD

The final step is to remaster our running real-time system using Remastersys package. Before we move to the explanation of this stage, we give a brief description of this tool. In fact, it's a Debian oriented remastering tool, controllable using command line or Graphic User Interface. It enables the creation of live installable CDs or DVDs, including all the software available in the installed system. We can choose to include or not our personal data by choosing between "dist" and "backup" parameters. We must add Remastersys repository, install and use it to remaster our system to obtain an ".iso" burnable image usable as live installable DVD. This phase is the easiest step in the realization thanks to the simplicity of Remastersys usage and the wide choice of parameters offered by this package.

### D. Testing the real-time characteristics of our system

The realized system can be used as live DVD or installed in a standard PC architecture. The real-time performances may vary depending on the used architecture. To have a clear idea about reached performances by deployment platform, Xenomai offers a set of benchmarks able to test different real-time aspects of the system. The most important benchmarks are described in Table II.

TABLE II. XENOMAI ASSOCIATED BENCHMARKS

| Benchmark | Description |
| --- | --- |
| Switchtest | Can test thread context switches. |
| Switchbench | Can measure the contexts switch latency between two real-time tasks. |
| Cyclictest | Can be used to compare configured timer expiration and actual expire time. |
| Clocktest | Can be used to repeatedly prints a time offset compared to reference gettimeofday(). |

These benchmarks can be used to familiarize students with real-time performance evaluation and their different associated metrics. Such can be illustrated by the evaluation of the impact of real-time enhancements into the overall system performances.

## V. CONCLUSION

The realized live/installable DVD can be used both in education or system development. The main contribution of such solution is to have a ready to run system, which minimizes the time of selecting and including different needed software components. This system can be enhanced and remastered after its installation and can be tuned by inclusion of new components to meet specific application needs.

This kind of solution offers the possibility to work with real-time Linux without losing the contact with classic RTOS knowledge's. It can be a very interesting way to introduce real-time and embedded Linux word especially when considering that Xenomai is actually used by various companies such as Sysgo in their ELinOS solution.

Considering that this distribution does not take the advantage of the two other predominant real-time Linux extensions (RTAI and RT-PREEMPT). We plan to extend our distribution with these two extensions by including multi configuration boot capability, which can allow the user to choose between these three alternatives. In the other hand, we plan to include and explore other open-source components that can be used for real-time applications design and code generation such as Topcased, Openembedd and Beremiz.

## AUTHORS PROFILE

N. LITAYEM received the Dipl.Ing. and M.S. degrees in electrical engineering from National School of Engineer of Sfax (ENIS), Tunisia, in 2005 and 2009, respectively. He received the MS. degree in Embedded Systems Engineering from the National Institute of Applied Science and Technologies, Tunisia in 2009. Currently, he is a Ph.D student with the "Laboratoire d'Etude et de Commande Automatique de Processus" (LECAP) at the university of Carthage (INSAT-EPT). His research interests are the reliable control of electrical drives using FPGA technologies.

A. BEN ACHBALLAH received the BSc degree in Electronics from Bizerte's Faculty of Sciences in 2007 and the MSc degree in Instrumentation and Measure from the National Institute of Applied Sciences and Technology of Tunis (INSAT) in 2009. Currently, he is a PhD Student with the "Laboratoire d'Etude et de Commande Automatique de Processus" (LECAP) at the Polytechnic School of Tunisia (EPT). His research interests include FPGA-based simulators for embedded control applications, simulation methodologies for network-on-chips and high level synthesis technique.

S. BEN SAOUD (1969) received the electrical engineer degree from the High National School of Electrical Engineering of Toulouse/France (ENSEEIHT) in 1993 and the PhD degree from the National Polytechnic Institute of Toulouse (INPT) in 1996. He joined the department of Electrical Engineering at the National Institute of Applied Sciences and Technology of Tunis (INSAT) in 1997 as an Assistant Professor. He is now Professor and the Leader of the "Embedded Systems Design Group" at INSAT - University of Carthage. His research interests include Embedded Systems Architectures, real-time solutions and applications to the Co-Design of digital control systems and SpaceWire modules.